# A Cloud-connected NO$_2$ and Ozone Sensor System for Personalized Pediatric Asthma Research and Management

Quan Dong, Baichen Li, R. Scott Downen, Nam Tran, Elizabeth Chorvinsky, Dinesh K. Pillai, Mona E. Zaghloul, *Fellow, IEEE*, and Zhenyu Li[*], *Member, IEEE*

*Abstract*—This paper presents a cloud-connected indoor air quality sensor system that can be deployed to patients' homes to study personal microenvironmental exposure for asthma research and management. The system consists of multiple compact sensor units that can measure residential NO$_2$, ozone, humidity, and temperature at one-minute resolution and a cloud-based informatic system that acquires, stores, and visualizes the microenvironmental data in real-time. The sensor hardware can measure NO$_2$ as low as 10 ppb and ozone at 15 ppb. The cloud informatic system is implemented using open-source software on Amazon Web Service for easy deployment and scalability. This system was successfully deployed to pediatric asthma patients' homes in a pilot study. In this study, we discovered that some families can have short-term NO$_2$ exposure higher than EPA's one-hour exposure limit (100 ppb), and NO$_2$ micro-pollution episodes often arise from natural gas appliance usage such as gas stove burning during cooking. By combining the personalized air pollutant exposure measurements with the physiological responses from a patient diary and medical record, this system can enable novel asthma research and personalized asthma management.

*Index Terms*—Cloud informatic system, indoor air quality, NO$_2$, Ozone, pediatric asthma research and management, personal exposure sensor

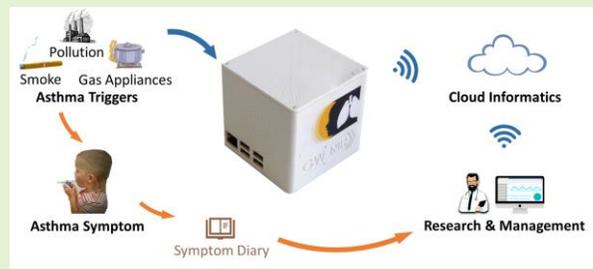

## I. Introduction

Asthma is the leading chronic disease in children [1], with more than 6.2 million pediatric patients (8.4% of child population) in the US alone [1], [2]. Exacerbations of asthma symptoms can be life-threatening and lead to approximately 629,000 emergency department visits and 13.8 million missed school days every year [3]. The annual healthcare and economic costs due to pediatric asthma have reached $5.9 billion in 2012 [4].

Although studies have identified many asthma triggers such as air pollution, allergies, respiratory infections, etc. [5], the relationship between triggers and exacerbations in terms of susceptibility and timing is not well understood. This is partly due to the lack of effective tools to accurately quantify personal exposure to these triggers [6]. In current clinical practice and academic research, the most common tools for exposure estimation are questionnaires, data from public sensor networks such as the Environmental Protection Agency (EPA) sensor network, and indirect estimations such as proximity to a major road. However, the data from questionnaires are very subjective and cannot provide quantitative measurements such as concentration and duration of air pollutant exposure. Public sensor networks are often limited by the number and locations of sensor stations, which makes it almost impossible to obtain true personal exposure profiles. Moreover, the time resolution of these pubic data is typically very coarse. For example, the data available from the EPA are provided as hourly averages and 8-hour averages. Such low time resolution is likely to miss short-term trigger exposure episodes such as cigarette smoke. Besides, most environmental sensor networks only monitor outdoor air quality whereas both adults and children spend the majority of their time indoors [7], [8]. To address these issues, a personalized sensor that can continuously monitor indoor exposure with a high temporal resolution is needed. In addition, a cloud-based informatic system would likely be required to accommodate a large number of sensors and human subjects in large scale epidemiological studies.

Manuscript received April 16, 2020. This work was supported by National Institute of Biomedical Imaging And Bioengineering (NIBIB) of the National Institutes of Health (NIH) under Award Number 1U01EB021986-01.

Q. Dong, B. Li, R. S. Downen, and N. Tran are with the Department of Biomedical Engineering, The George Washington University, Washington, DC 20052 USA (e-mail: dongquan@gwu.edu; baichen@gwu.edu; sdownen@gwu.edu; nam_tran@gwu.edu).
E. Chorvinsky is with Division of Pulmonary Medicine, Children's National Hospital, Washington, DC 20010 USA (e-mail: EWILLIAMS3@childrensnational.org).
D. K. Pillai are with the Division of Pulmonary Medicine, Children's National Hospital, Washington, DC 20010 USA, and is also with the Departments of Pediatrics and Genomics & Precision Medicine, the George Washington University, DC 20052 USA (e-mail: DPillai@childrensnational.org).
M. E. Zaghloul are with the Department of Electrical Engineering and Computer Engineering, The George Washington University, Washington, DC 20052 USA (e-mail: zaghloul@gwu.edu).
*Z. Li is with the Department of Biomedical Engineering, The George Washington University, Washington, DC 20052 USA (e-mail: zhenyu@gwu.edu).

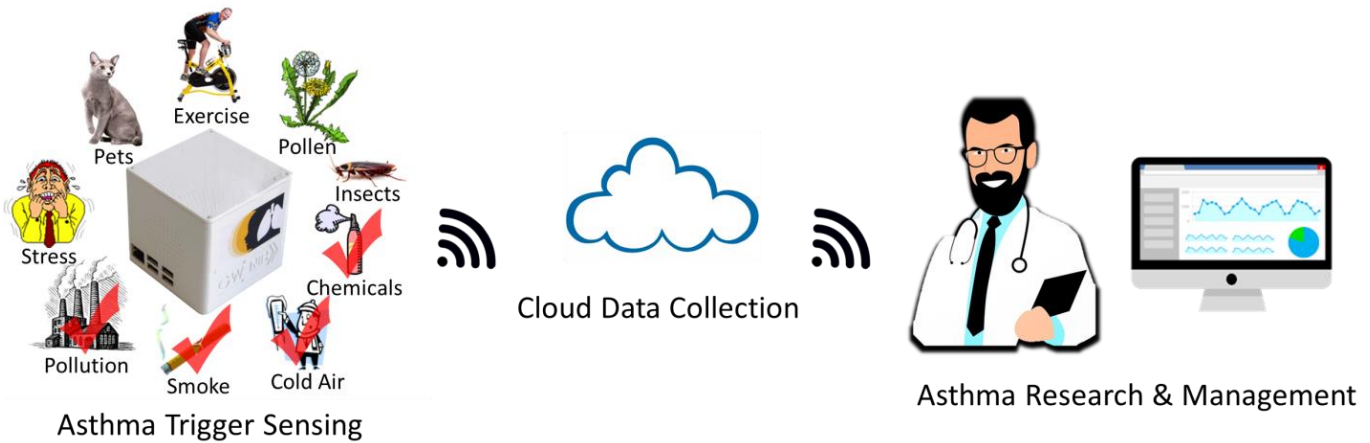

Fig. 1. A personalized asthma trigger sensor system with cloud connectivity. The sensor system contains one or multiple compact sensor units designed to measure personal asthma trigger exposure. The current sensors focus on air pollutants from factories, smoke, or some indoor appliances, and some physical measurements including temperature (cold air) and humidity. All the sensors are connected to a cloud based informatic system currently implemented on AWS. Asthma trigger exposure data will be collected and stored on the cloud informatic system, and researchers and doctors can access and visualize these data online or download the data to local computers for further analysis.

Among all asthma triggers, air pollution is one of the most common, affecting 91% of the world's population according to World Health Organization (WHO) [9]. $NO_2$ and ozone are among the top six most common air pollutants regulated and monitored by the EPA as a requirement by the Clean Air Act [10]. Both $NO_2$ and ozone are known airway irritants and asthma triggers [5]. Common sources of $NO_2$ include automobiles, gas cooking and heating appliances, and environmental tobacco smoke (ETS) [11], [12]. Speizer et al. found that the $NO_2$ level measured one meter away from the gas oven can be constantly above 200 parts per billion (ppb) for more than one hour while the appliance is on, and that the peak valve can reach values above 550 ppb [13]. This exposure is far above the EPA's one-hour exposure limit (100 ppb). A long-term cohort study warns that $NO_2$ is likely to increase the prevalence and severity of asthma in children [14]. Another recent large-scale epidemiological study estimates that more than 90% of $NO_2$ attributable pediatric asthma incidences occurred in areas with annual average $NO_2$ concentrations below 21 ppb, the WHO annual average limit [15]. However, some human subject experiments have found little to no observable respiratory response associated with short-term $NO_2$ exposure, even those including asthmatic patients exposed to $NO_2$ at a level of 4 parts per million (4000 ppb) [16]–[19]. The mechanism behind $NO_2$ exposure and asthma remains unclear in the literature and the consensus threshold level (concentration and duration) at which $NO_2$ exposure causes adverse health effects is controversial [15], [19], [20]. To resolve the inconsistency between the long-term large-scale epidemiological studies and short-term exposure studies, easy-to-deploy personalized $NO_2$ sensors are needed.

Ozone is another known asthma trigger, which is usually a secondary and transient air pollutant generated by chemical reactions between oxides of nitrogen (NOx) and volatile organic compounds (VOC) in the presence of sunlight [21]. A recent systematic review of outdoor air pollution and asthma in children concluded that chronic exposure to ozone may increase the risk of hospital admissions due to asthma among children [22]. Therefore, it is worthwhile to also study indoor ozone micropollution and its effects on pediatric asthma patients.

In this work, we propose a cloud-connected $NO_2$ and ozone sensor system that can measure personal indoor pollutant exposure at high temporal resolution. The sensor utilized in this study has a lower limit of detection (LLOD) of 10 ppb for $NO_2$, whereas the EPA's annual limit is 53 ppb [23], [24]. The sensor's LLOD for ozone is 15 ppb, which is well below the EPA annual limit of 70 ppb [25]. The sensor measures $NO_2$ and ozone along with temperature and humidity every minute to ensure the capture of transient episodes of exposure such as cigarette smoking or cooking with a gas-powered appliance. The cloud-based informatic system can facilitate data collection for a large population study. Additionally, doctors and researchers can use this informatic system to view the real-time exposure (given that the privacy and safety issues are addressed) in order to provide timely instructions and feedback to the patients for personalized asthma management and treatment in the future.

## II. MATERIALS AND METHODS

### A. A personalized asthma trigger sensor network with cloud functionalities

As shown in Fig. 1, the complete system consists of: multiple compact personalized asthma trigger sensor units distributed at multiple locations, a cloud-based informatic system, and a web-based graphical user interface (GUI) for physicians and researchers. The current sensor unit can measure gas pollutants ($NO_2$ and ozone), temperature, and humidity. It is implemented using a modular design approach. Inside the sensor unit, a custom printed circuit board (PCB) is used to interface with commercially available gas sensors. The sensing capabilities can be expanded by integrating more sensors such as $SO_2$ or PM sensors into the customized interface board, depending on the application.

The sensor unit is 9.1 cm × 9.1 cm × 9.0 cm (L × W × H) and weighs 214 g. It is designed for indoor use such as at home, school or in the workspace, and starts the data acquisition process automatically on power up, allowing the sensor to be effortlessly deployed with minimum maintenance. The sensor unit has built-in Bluetooth, Wi-Fi, and Ethernet connection capabilities. When Internet is available, it will upload acquired data to the cloud-based informatic system. Additionally, the sensor unit can also connect to a gateway device (such as a smartphone) via Bluetooth, which can then relay the sensor data to the cloud informatic system.

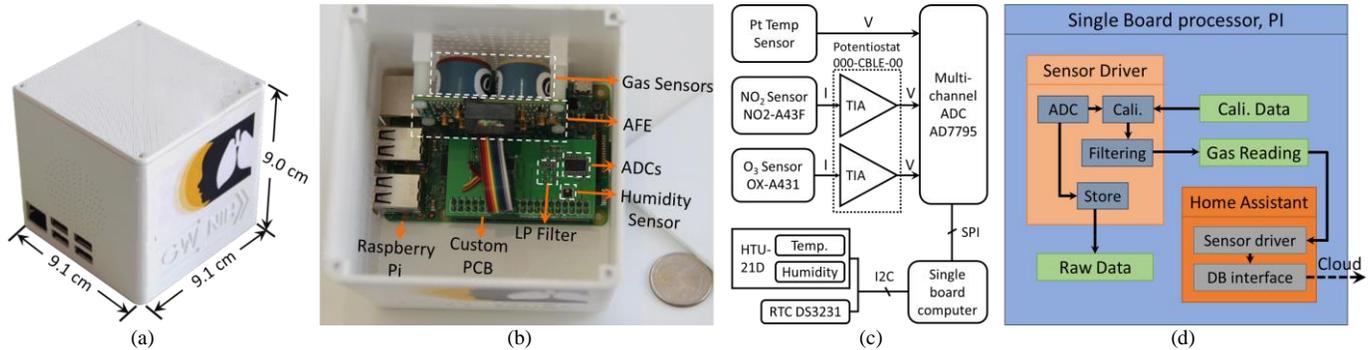

Fig. 2. Compact gas sensors for indoor application. A) photograph of the sensor with dimension labeled; b) photograph of the internal structure of the sensor when the top cover is removed; c) block diagram of the hardware structure of the sensor device, which includes sensors, custom data acquisition PCB and a single board computer Raspberry Pi; d) the block diagram of the software structure of the Raspberry Pi, which includes the sensor driver, calibration setting, an open-source software, Home Assistant, database and internet communication handling, and a raw data storage.

The cloud informatic system forms a data communication pathway between the sensors and the researchers/physicians by aggregating sensor data automatically through the Internet connection and displays it to the researchers/physicians through a web-based interface. This system is particularly helpful to reduce the workload for data collection when a large number of sensors are deployed. The cloud system includes two major parts: a database and a data visualization web server. The database collects and stores the data measured by the sensors while the data visualization server displays these measurements in real-time along with historical data utilizing a set of graphing tools including time-series plots, bar charts, histograms, etc. In addition, raw data can be downloaded to a local computer through the web-based user interface for further analysis.

The sensor unit can also synchronize data from the cloud server to its local storage for bidirectional communication, control, and feedback, which provides the potential for remote care of patient in the future. When the Internet is not available, the sensor unit can function independently by saving data locally to a 32GB micro SD card, which is sufficient for more than five years of continuous operation.

In our current setup, the default sampling rate is one sample per minute for all measurements, including temperature and humidity. This sampling rate is sufficient for most transient events such as cigarette smoke or cooking with a gas-powered appliance. A raw measurement is taken every two seconds and every 30 data points are averaged into one recorded sample to reduce noise and improve accuracy. All the raw data before averaging are stored locally on the SD card for debugging purposes in case an unexpected event happens.

### B. Compact gas sensor unit for indoor application.

The compact gas sensor unit is approximately a 9 cm cube and weighs 214g, as shown in Fig. 2a. It consists of: 1) two commercially available electrochemical $NO_2$ and ozone sensors, 2) a custom-built data acquisition PCB for data collection, 3) a single board controller based on Raspberry Pi [26] as shown in Fig. 2b,c. Below we will describe these three components in detail.

1) *Electrochemical NO2 and Ozone sensors*

There are a variety of sensing modalities available for gas sensing, such as optical spectroscopy [27], gas chromatography [28], chemiluminescence [29], ion-selective electrodes [30], etc. However, most of them are too complex and prohibitively large for a home setting. Metal oxide CMOS sensors are compact and require minimum maintenance. Unfortunately, a majority of metal oxide sensors only measure gas concentration around the ppm level and with poor selectivity [31]–[33]. Therefore, the electrochemical sensor was chosen as a compromise between size and performance.

The two gas sensor elements shown in Fig. 2b are manufactured by Alphasense [34]. The $NO_2$ sensor, NO2-A43F, can detect $NO_2$ as low as 10 ppb. The ozone sensor, OX-A431, has an LLOD of 15 ppb. These two sensors both have a cylindrical shape of $\Phi 20.2$ cm × 16.5 cm and weigh 5g.

The electrochemical sensor separates the reduction and oxidation (redox) reactions onto two electrodes with the help of an external circuit, and electrical current generated during the redox reaction will flow through the external circuit. In the case of $NO_2$, the reduction happens on an electrode called the working electrode, which will reduce $NO_2$ into NO and consumes two electrons [35], [36],

$$NO_2+2H^++2e^-\rightarrow NO+H_2O \qquad (1)$$

Current generated by the redox reaction is proportional to the reaction rate of the target reactant. For a properly designed sensor, the reaction rate is controlled by the ambient target concentration in a linear relationship. Overall, an electrochemical sensor will output a current signal proportional to the target gas concentration, and the proportional coefficient, i.e. the responsivity, should be constant under a given working condition. After the responsivity is obtained though calibration or factory specifications, the target gas concentration can be calculated with the responsivity and the current readout.

Typically, the electrochemical sensor (cell) consists of 3 electrodes: a working electrode (WE), a reference electrode (RE), and a counter electrode (CE). The WE is where the target gas reaction takes place. The electrical potential of the WE will affect the reaction rate. As a result, an external circuit, called a "potentiostat," is usually implemented to stabilize the WE voltage (bias) with respect to the RE by varying the voltage of the CE. The RE provides an accurate reference point for the electrochemical cell. The potentiostat also works as a transimpedance amplifier (TIA) by converting the reaction current to a measurable voltage output.

There are several electroanalytical methods which can be used with the electrochemical cell and potentiostat setup, including amperometry, voltammetry, coulometry, etc. [37]. In this work, the amperometry method [38] is used, and it is also the most widely used method in this gas sensing scenario. The amperometry setup biases the WE at the optimal potential for the target gas and measures the current generated by the target gas reaction. The WE bias thermodynamically favors the redox reaction for the target analyte to increase selectivity. Without the presence of any interference, the current is proportional only to the target gas concentration.

For the amperometry setup, the sensor unit utilizes a potentiostat board, 000-CBLE-00, from Alphasense. This board is also referred to as an analog front end (AFE), as it converts the original measurement to a voltage output that can be digitalized by an analog to digital converter (ADC). A high accuracy platinum temperature sensor is located onboard adjacent to the two sensors. By measuring the temperature, temperature-induced drift can be compensated for in software.

As with most sensors, the electrochemical gas sensor produces a baseline current, i.e. zero current, even when there is no reaction of the target gas. Zero current compensation or removal is essential for correct gas concentration calculation, especially at low concentration levels. However, the removal is not straightforward, as zero current drifts not only with the temperature changes but also with other factors such as humidity changes [39] and sensor degradation over time [40]. In our design, the sensing elements have a duplicated working electrode, which is called an auxiliary electrode (AE). The AE is of the same construct as the WE and in the same environment but is contained inside without exposure to the target gas. As a result, the total current generated by the AE will approximate the zero-current of the WE at different temperatures or in the presence of other interference. As a result, the compensated WE current, WEc, can be calculated using the following equation:

$$WEc = WEu - n_T \times AEu \qquad (2)$$

WEu and AEu are the uncompensated currents generated by WE and AE, respectively. The term $n_T$ is a temperature correction coefficient provided by the vendor. $n_T$ compensates for the drift variance of AE and WE at different temperatures and is very close to 1. After the zero-current compensation, we also performed additional in-house calibration with a gas standard generator. The calibration parameters are saved on the Raspberry Pi to further compensate for the zero current and responsivity variations for the individual sensor. More details will be discussed in the calibration section.

2) *Custom data acquisition PCB*

A custom PCB provides an interface to collect all the measurement data, including pollution gases, humidity, temperature, and timestamps as shown in Fig. 2b,c. The onboard high-resolution low-noise ADC, AD7795, digitizes the voltages from the gas sensors and Platinum (Pt) temperature sensors. It has an Effective Number of Bits (ENOB) of 16, which provides a sufficient dynamic range for the indoor gas sensing application, and has eight input channels, which could be used for future sensor expansion. The ADC communicates with the Raspberry Pi through a Serial Peripheral Interface (SPI) bus. This SPI bus provides a convenient way to adapt additional ADC chips when more than eight channels are needed. In front of the ADC, passive low-pass RC filters with a cutoff frequency of 16 Hz are implemented, which will remove any unwanted high-frequency noise, interference, and aliasing. The internal 60 Hz digital filter is enabled to further reject the power line interference in the home setting.

In additional to performing Analog-to-Digital conversions, the interface board also measures ambient humidity from a CMOS sensor (HTU-21D) and maintains time using a high accuracy real-time clock (RTC) chip (DS3231). These two chips are on an Inter-Integrated Circuit ($I^2C$) bus. Humidity fluctuations, besides being one potential asthma trigger [41], are also a potential source of sensor drifts, thus its measurement may be needed for a more sophisticated sensor compensation algorithm [39]. The RTC maintains the accurate time after the Raspberry Pi is powered off or when the Internet time is not available. The sampling rate is controlled by the driver software on the Raspberry Pi and can be changed as needed. The driver also instructs the ADC chip to perform self-recalibration using its internal reference periodically to correct the offset drift and gain drift caused by the electronics.

3) *Single board controller based on Raspberry Pi*

A single board controller based on the Raspberry Pi 3 is implemented in the sensor unit to handle the sensor control, measurement acquisition, data storage, and cloud connection, as shown in Fig. 2d. Raspberry Pi 3 is a low-cost single-board computer with a 1.2 GHz ARM V8 processor core and a variety of connectivity options, including Ethernet, Wi-Fi, Bluetooth 4.0, SPI, and $I^2C$. The operating system (OS) running on the Pi is Raspbian, an ARM compatible Linux distribution [42]. Sophisticated algorithms and software can be deployed quickly and easily with the help of many open-source packages and information from the free and open-source community.

To ease the software development procedure, an open-source home automation software platform, Home Assistant (HA) [43], is installed on the Raspberry Pi. HA provides an automation framework with many available components, including a sensor control framework, local databases, a web-based remote database interface, etc. The sensor framework communicates with the sensor driver to fetch the sensor reading, and the local database buffers the readings temporarily before sending them to the remote database. When the Internet is available, the remote database will be linked based on the username, password, and credential settings and upload the sensor readings. HA also provides a web-based GUI. When authorized, patients can view their exposure data through this GUI. This web interface can be accessed from any device with an Internet browser installed, including PCs and smartphones, which eliminates the need for developing multiple native apps for different device platforms. HA can also run custom add-ons, which can be quickly developed using the Python programming language. Custom HA add-ons give us the ability to control all hardware and software resources available to HA through the web interface, such as updating the sampling rate. With the help of the web interface and a custom add-on, we implemented a Wi-Fi setup function to help the user connect to their home Wi-Fi when the research team is not available.

The device driver for the custom PCB interface board is written using the C programming language. The driver performs ADC sampling, calculates gas concentrations based on the temperature and stored calibration data, filters the noise digitally, and stores the reading on-board, as shown in Fig. 2d. The driver auto-starts as a daemon when the Linux OS starts and driver communicates with the ADC through an SPI bus. The AE and WE voltages are read every 2 seconds for both gas sensors, and the voltage of the Pt sensor is read every minute. Then the equation (2) is applied to minimize the zero current. After that, the calibration data are loaded, and the voltage is converted to the gas concentration. The calculated concentration is filtered and is down-sampled to the desired sampling rate as the final gas concentration reading, which can be read by HA. This process decreases the noise and reduces the data load for the local storage, the Internet, and the server. In addition, the driver also reads the humidity every minute from the CMOS humidity senor on the interface board through an $I^2C$ interface. All the raw and derived data can be stored on the SD card for debug purposes. The processed concentration, temperature, and humidity are accessible by HA for local visualization and are relayed to the server by HA.

The timestamp is an essential part of the sensor measurement data. The system clock is maintained both by the Network Time Protocol (NTP) and a hardware real-time clock (RTC). When the device is connected to the Internet, the system clock is synchronized by the NTP, just like in a typical PC and smartphone. However, the Raspberry Pi itself cannot maintain its clock when it loses power as it does not have an always-on timer. To address this issue, a high-accuracy RTC is added on the interface board and runs on a CR1220 battery. It drifts less than 2 minutes per year from -40°C to +85°C and maintains the clock even when Raspberry Pi loses power. When the Internet is available, Raspberry Pi synchronizes the RTC with the NTP periodically to maintain its accuracy.

### C. Cloud-based informatic system

The cloud-based server was implemented in order to facilitate data aggregation and visualization, as shown in Fig. 3, when a large number of sensor units are deployed in the field. It collects both the indoor environmental data measured by the compact sensor units and the outdoor data provided by the EPA sensor network. All the collected data can be viewed and downloaded by researchers and physicians though the HTTP web interface using a web browser.

An Ubuntu based Linux server is hosted on the Amazon Elastic Compute Cloud (Amazon EC2) provided by Amazon Web Service (AWS) [44]. Two open-source packages, InfluxDB [45] and Grafana [46] are installed to handle the time series database functions and data visualization, respectively.

The InfluxDB 1.6.4 was installed on the cloud server and the built-in authentication based on user credentials was manually enabled. As shown in Fig. 3, two separate databases were created to store the compact sensor data and the EPA data. Each measurement is stored as a data entry with a timestamp, the serial number of the sensor unit, measurement type, measurement value, and the unit of the measurement. As a NoSQL database, the format of each data entry is not restricted to a predefined schema. This simplifies the process of expanding or modifying the database, as any arbitrary type of measurement can be added freely, such as a new gas type or physiological measurements from a patient. Besides actual measurements, the sensor metadata, which stores sensor characteristics, status, etc., can also be attached the current sensor data entries in the future without a redesign of the current database. All data storage and retrieval of information is performed through the HTTP interface of InfluxDB.

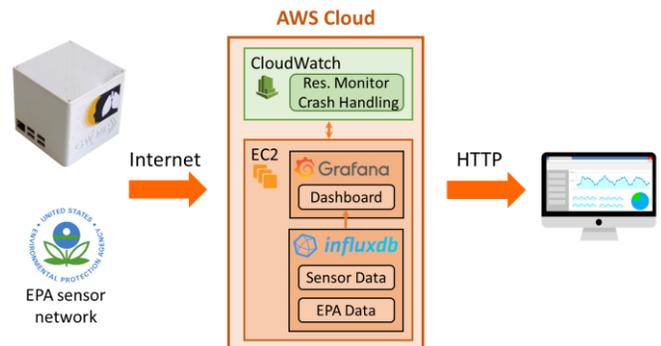

Fig. 3. Cloud-based informatic system. The cloud system facilitates data acquisition by collecting and storing data from the sensor units automatically with the help of a time-series database, InfluxDB. A protected data visualization server which researchers or doctors can log in to allows viewing data in real time, simple analysis, or downloads of data to a local computer for further analysis. Besides collecting data directly from sensors inside our own sensor networks, the cloud system has APIs to retrieve data from 3rd party sensor networks, such as EPA sensor networks.

The EPA outdoor air quality data from the DC/Maryland/Virginia (DMV) region comprising the area of our current study cohort are acquired via the AirNow application programming interface (API) [47] and stored in InfluxDB using the official InfluxDB Python package. The API returns air quality measurements collected from multiple stations near a particular zip code in JSON format through an HTTP GET request. These requests are made every half hour through a scheduled AWS Lambda function implemented in Python using the requests Python package. AWS Lambda is a serverless computing platform that runs a function when needed without constantly running a server. As a result, it offloads the request and store operation from our cloud server.

For a user interface, we configured Grafana, an open-source data analytics and interactive visualization software with a web interface provided by its internal webserver, to display current and historical sensor data. The data are retrieved from InfluxDB and organized by patient ID. A patient can simply view his/her own data, while researchers with higher privileges can customize the dashboard, perform simple data analytics, set up alerts, and download the historical data to the local computer for further analysis. Grafana version 5.3.2. was installed and configured on the cloud server in a similar process as described for InfluxDB.

Besides the EC2 instance that runs the Linux system, AWS also provides cloud monitoring and management utilities such as CloudWatch. CloudWatch monitors resource usage such as CPU, RAM, hard drive, and Internet traffics. A key advantage of a cloud-based system is scalability. The CloudWatch resource usage monitoring allows the system to automatically scale up the cloud server with a service upgrade when needed. There is no need for the user to maintain the hardware. Moreover, the specifications of the virtual server can be upgraded anytime by choosing a different tier of service. In addition to resource monitoring, the CloudWatch also performs crash handing. When a system crash is detected, the CloudWatch reboots the server to minimize the system downtime. Higher tiers of service also include load balancing with multiple servers in multiple regions, which can be useful for large-scale population studies. Besides load balancing, multi-server setup also increases redundancy, which can reduce system downtime and latency for multi-region studies by allowing the sensor units and researchers to the closest server instance.

## III. Results

To ensure accuracy, the sensor was first calibrated and characterized in a lab environment. After the lab calibration, the sensor was tested in a real-life apartment setting to validate the system performance and reliability. Finally, a pilot study was carried out by deploying the sensors into pediatric asthma patients' homes to monitor microenvironment pollution.

### A. Sensor calibration with permeation tube

Typically, each individual sensor has small performance variations, including responsivity and offset. Although the manufacturer provides a factory calibration, sensor performance might drift during the process of shipment and sensor unit fabrication. As a small drift can cause a large concentration variation in the ppb level measured, after the sensor unit was assembled, we recalibrated the responsivity and offset rigorously in the lab using the setup shown in Fig. 4. The calibration data for each sensor was then saved on the SD card of the Raspberry Pi controller for the gas concentration calculation.

The calibration setup starts with a permeation tube gas standard generator, FlexStream™ Base Module, from KIN-TEK Analytical Inc, as shown in Fig. 4a,b. The permeation tube is a small tube of chemical encapsulated in a permeable membrane, and it will dispense a constant amount of chemical vapor at a constant temperature. The gas standard generator has an internal oven and can heat the permeation tube at a precise and stable temperature to generate the target gas at a steady rate. After the target gas is generated, the generator will mix it with a precise amount of diluent gas in order to achieve a specified concentration using the internal mass flow controller. The $NO_2$ permeation tube is from KIN-TEK and certified at 50 °C with an emission rate of 187 ng/min. This tube and generator combination can generate $NO_2$ from 18.2 ppb to 364 ppb. We set the gas concentration at 20, 50, 100, 150, and 200 ppb for the sensor calibration.

The gas standard generator in this work controls the gas concentration by varying the flow rate. As a result, the flow rate of the target gas varies greatly at the set calibration concentrations. To eliminate the variation of the total flow rate on the sensor surface, we used a back-pressure regulator and an orifice in front of the gas chamber. The orifice is a flow regulator with a small hole along the flow path. The pressure difference between its two sides controls the flow rate. The output side of the orifice is connected to the gas chamber in which the pressure is close to that of the atmosphere due to the low flow resistance of the chamber. The input pressure of the orifice is controlled by the back-pressure regulator, which releases the excess gas to the vent port and maintains the pressure at the orifice input. The orifice used is a push-to-connect flow-control orifice, 6349T42, with 0.01" orifice diameter from McMaster-Carr. The orifice input pressure is set at one psi and can be monitored by the pressure gauge.

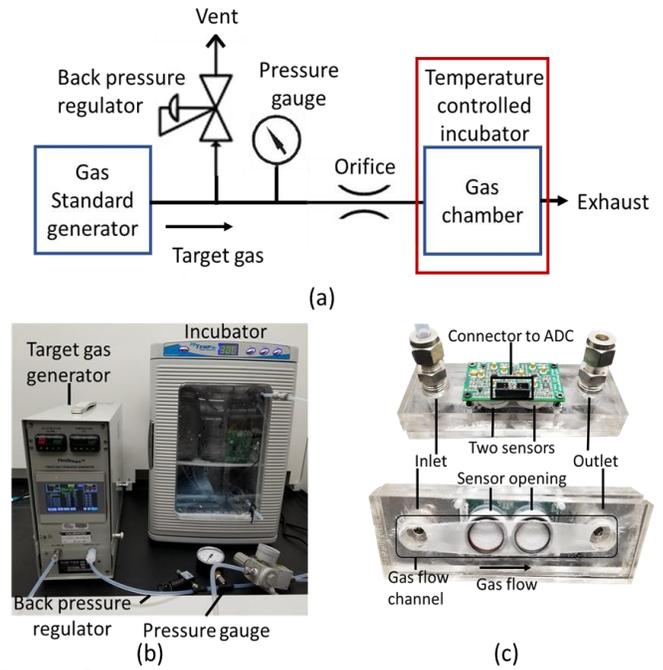

Fig. 4. Sensor calibration setup with a permeation tube based gas standard generator: a) block diagram of the gas flow setup where the gas generator generates the target gas at predefined concentrations with a constant flow rate. The orifice and back pressure regulator maintain the flow rate into the gas chamber by venting the excess gas. The gas chamber hosts our sensors inside. The incubator regulates the temperature of the gas chamber at set values; b) photograph of the physical setup; c) the gas chamber design, the sensor is mounted on top of the flow channel, where the target gas flows.

The gas chamber exposes the sensor to the generated target gas though an airtight interface, as shown in Fig. 4c. The chamber is made of high-density acrylic with an inlet and an outlet connected by a gas flow channel. In the middle of the flow channel, the two sensors are mounted to the two sensor mounting openings with the sensor apertures facing inwards. The interface between the chamber opening and the sensor housing is sealed with two O-rings to prevent leakage. The gas chamber is placed in a temperature-controlled incubator to verify the temperature dependence of the sensor responsivity and offset. As the sensor unit is designed for indoor application, the test temperature is set at 15ºC, 20ºC, 30ºC, and 40ºC.

After the calibration, all the characteristics are stored on the sensor unit, and the sensor unit displays the gas concentration in ppb. To validate the performance of the calibrated sensor units, a series of $NO_2$ concentrations are flowed on to the sensor units using the same setup as in Fig 4. As shown in Fig. 5a, 200, 150, 100, 50, and 20 ppb of $NO_2$ are supplied to the sensor from highest to lowest concentration. Each gas concentration is held for 30 minutes before being purged with zero air (filtered air with no target gas). The sensor can clearly measure $NO_2$ concentration at 20 ppb with low hysteresis and quick response. The noise equivalent concentration is 10 ppb. Fig. 5b shows the linearity and the offset of the sensor by plotting the sensor reading on the y-axis and the generated gas standard on the x-axis. The sensor reproducibility is validated by alternating the $NO_2$ at 100 ppb, and zero air every 30 minutes repeatedly, as shown in Fig. 5c. The sensor reading fits along the 100-ppb line when the gas is on and along the 0-ppb line when the zero air is on. In this test, the sensor drift is not apparent.

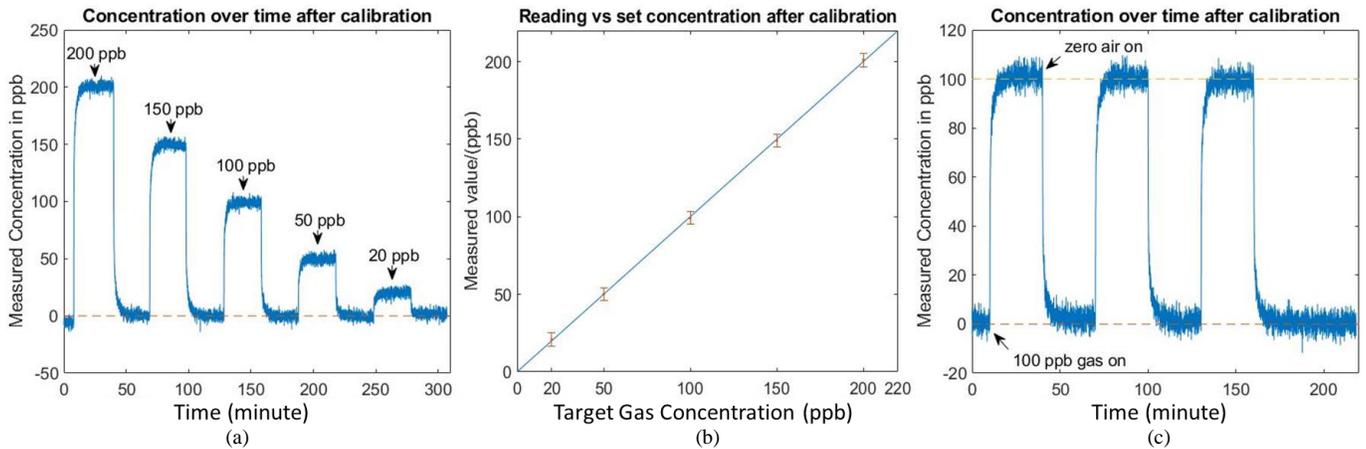

Fig. 5. Sensor calibration results: a) the sensor response to different concentrations of $NO_2$, at 200 ppb, 150 ppb, 100 ppb 50 ppb 20 ppb with low hysteresis and a LLOD of 10 ppb; b) the senor readings on the y axis with respect to the $NO_2$ concentrations generated by the gas generator show good linearity and zero offset; c) repeated on-off test to ensure baseline drift is not an issue.

The ozone sensor was used as manufactured without further calibration, and we did not observe significant ozone fluctuations recorded throughout our asthma trigger study inside pediatric asthma patients' homes.

### B. Sensor validation in the home setting

To ensure the sensor unit can function reliably in a real-world environment, we validated our sensors in an apartment home after the lab calibration before deploying them into the patients' homes.

According to the literature [11], [12], natural gas stoves are a primary source of $NO_2$. We first tested the sensor response to a gas stove burning in a one-bedroom, one-bathroom (1b1b) apartment with an open kitchen setting. The sensor was placed in the living room about 7 meters away from the stove. $NO_2$ concentration was highly associated with gas stove usage, as shown in Fig. 6a. The $NO_2$ concentration rises rapidly after the stove is turned on and decreases slowly after the gas stove is turned off. The peak concentration observed was as high as 90 ppb. During this process, the sensor functioned normally without any observed error.

After this short period, single sensor test, we deployed three sensors into a 2-bedroom apartment with a gas stove for a week to test the intra-sensor variability and longtime stability. The sensors were placed in the kitchen, living, and bedroom. As shown in Fig. 6b, the $NO_2$ measurements from three sensor units in different rooms correlate with each other very well. Ozone is not shown in the figure as no ozone was detected by any of these three sensor units.

### C. Preliminary results from pediatric asthma patients' homes

After the sensors were validated in the lab and apartment setting, we conducted a pilot study on personal microenvironmental exposure in pediatric asthma patient homes, as shown in Fig. 7. The study was approved by the Institutional Review Board (IRB Pro00009593) of the Children's National Hospital (Washington, DC). All subjects were enrolled via signed informed consent from the Severe Asthma Program in the Division of Pulmonary and Sleep Medicine. Each consenting family received a sensor unit and a diary to collect patient symptom information and medication usage data. The sensor units were placed in the patient's home for six days, and families/subjects recorded symptoms in the diary during that time period. Both the device and diary were then mailed back to Children's National Hospital for data retrieval and analysis.

In our pilot study survey, we found that less than half of the patients in our cohort have Internet connections in their homes.

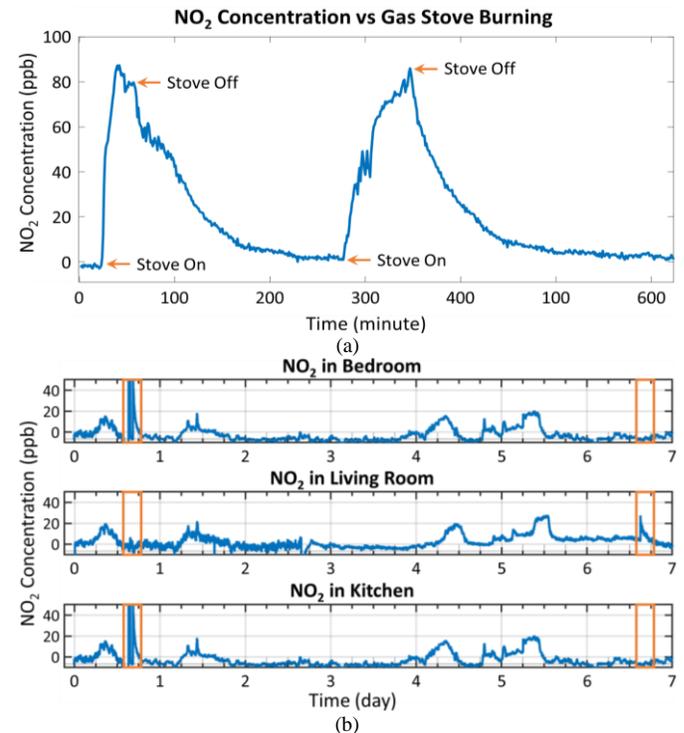

Fig. 6. Sensor validation in home setting: a) sensor response to gas stove in a 1-bedroom apartment with an open kitchen. The sensor was about 7m away from the stove. $NO_2$ level rises as the stove is on, peaks at 90 ppb and drops as the stove is off. b) Week-long sensor validation in a 2-bedroom apartment. Sensors were simultaneously deployed to the bedroom, living room, and kitchen. $NO_2$ levels in the three rooms correlate with each other strongly, despite some interferences on Day 1 and Day 7 denoted by the orange boxes.

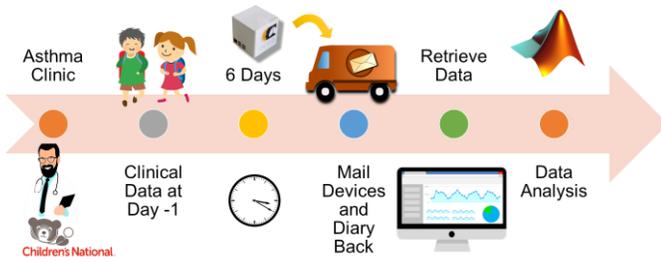

Fig. 7. Protocol for the pilot clinical study at Children's National Hospital in Washington, DC. Each dot on the arrow ribbon represents one step: 1) Recruitment, 2) Clinical data collection, 3) six days of deployment, 4) Device and diary retrieval, 5) Data retrieval, 6) Data analysis.

Thus, to avoid any potential bias, the sensor data were manually downloaded instead of using the cloud system. As expected, the manual data download increases the workload greatly and in turn reduces the deployment turnover rate.

Currently, we have deployed the sensors into over 30 pediatric asthma patients' homes. From the preliminary data, we observed that $NO_2$ exposure of families with gas stoves is higher than that of those without gas stoves. Fig. 8a is a measurement from a typical home without any gas appliance, "Home 1". $NO_2$ and ozone are effectively zero, and temperature and humidity are stable. In contrast, Fig. 8b is from a patient's home with a gas stove, "Home 2". Periodically high $NO_2$ episodes can be observed around lunch and dinner time, 12:00 and 18:00. The ozone, temperature, and humidity are not shown as they are very similar to these of Home 1. As seen in Fig. 8b, the $NO_2$ exposure has a very high temporal variation. Many of the details, such as peak exposure, will be lost if 1-hour and 8-hour averaging are performed on the exposure data from Home 2, as shown in Fig. 8c. Many outdoor sensor networks, such as the EPA's, only provide averaged data, which can be a limitation for epidemiological studies. The one-minute resolution of the sensor unit will preserve all the temporal information of the exposure, which can be useful for studies on the acute effect of the exposure or causal relationship between asthma symptoms and triggers.

## IV. DISCUSSION AND CONCLUSIONS

The sensor developed in this work provides nearly continuous air pollution monitoring of asthma trigger with a one-minute time resolution, which can be used to study potential exposure patterns that trigger asthma exacerbation and progression of disease severity attributable to air pollution gases. We have successfully deployed the sensors to pediatric asthma families for a minimum of six days, and up to two months. The sensors collected microenvironmental data reliably and showed that families with gas stoves in their homes can have significant acute exposure to $NO_2$ that may not be captured with other sensing methodologies.

However, challenges still remain. Our pilot study targets the inner-city pediatric population in Washington, D.C., and some low-income families in this populations may have limited Internet access. In order to avoid potential bias, we disabled our cloud system in this study at the expense of increasing the workload for data collection and reducing the turnover rate for sensor redeployment, due to the finite number of sensor units. We also encountered some user compliance issues, such as sensor misusage, that resulted in failed data collection and loss of sensors. Patients also did not reliably record symptoms in the diaries during the 6-day study period, prohibiting observations on the correlation between asthma symptoms and trigger exposures. In the future, the duration of individual deployments should be lengthened to increase the likelihood of observing the occurrence of symptoms. Additionally, more objective ways of capturing these symptoms are needed to effectively study the causal relationship between trigger exposure and asthma exacerbation, and the progression of disease severity.

## ACKNOWLEDGMENT

The authors thank the NIH PRISMS community, especially Dr. Katherine Sward, Dr. Neal Patwari, and Philip Lundrigan from University of Utah for helpful discussions. The content is solely the responsibility of the authors and does not necessarily represent the official views of the National Institutes of Health.

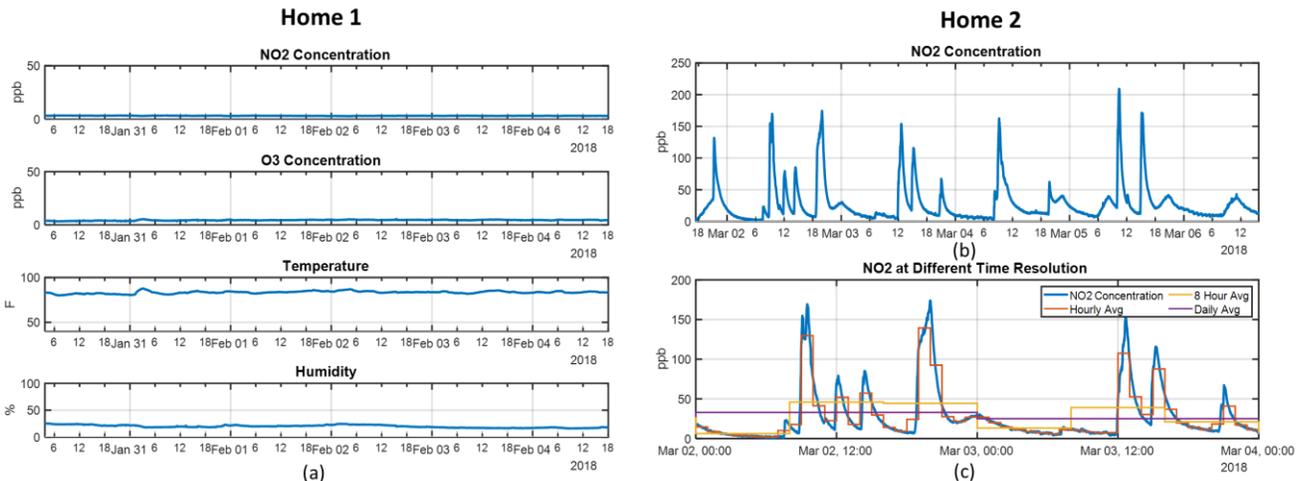

Fig. 8. Preliminary results from pediatric asthma patients' homes. a) Sensor measurement in Home 1 without any gas appliance. $NO_2$ and Ozone concentrations are effectively zero. The temperature and humidity are stable. b) $NO_2$ concentration in Home 2 with a gas stove. High $NO_2$ episodes occur every day around 12:00 (lunch time) and 18:00 (dinner time). Ozone, temperature and humidity are not shown as they are similar to the readings in Home 1. c) Loss of $NO_2$ fluctuation details with different averaging widows. Hourly average, 8-hour average, and daily average will smooth out the exposure details with longer average window retaining less details.

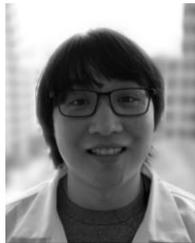

**Quan Dong** was born in Qingdao, China, in July 1988. He received a B.S. degree in Automation from the Beijing Institute of Technology, Beijing, China, in 2011.

Now he is a Research Assistant with the George Washington University (GWU), Washington, DC. His areas of interest include biomedical devices, wearable devices, and flexible electronics. He is currently working on a project related to biomedical devices for pediatric asthma research at GWU.

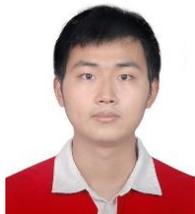

**Baichen Li** completed his master's degree in Biomedical Engineering at The George Washington University (GWU) in 2013. He worked in a medical device startup as an electrical engineer before he returned to pursue a Ph.D. degree in 2015 at GWU.

Currently, he is a PhD candidate in the Nanophotonics and Microfluidics Lab at GWU, conducting research on wearable sensors, wireless sensor network, point-of-care diagnostic devices, and microfluidics.

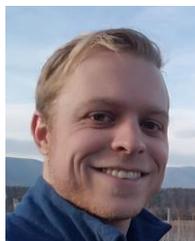

**R. Scott Downen** was born in Houston, Texas in 1988. He received a B.S. degree in Applied Sciences with a focus in Biomedical Engineering from the University of North Carolina, Chapel Hill, in 2011 and a M.S. degree in Biomedical Engineering from The George Washington University, Washington, DC, in 2017.

From 2011 to 2015, he worked as a design engineer and project manager for a medical device design consulting company based in Raleigh, North Carolina. Through this work, he holds a patent for a design for an automated electronic dose counter for inhalers. He then worked for a home security & home automation company based in the Washington, D.C. area as a design engineer. While there, he had a specific focus on design for manufacturing of mechanical and electrical designs of home automation, security, and sensing products. While pursuing his Master's Degree, his research involved computer vision and machine learning working towards the end goal of emotional state recognition from gesture analysis. Since 2018, he has been pursuing a Ph.D. in Biomedical Engineering with a focus in sensor networks, bio-instrumentation and low-power, wearable biosensors. His research interests include, but are not limited to, wearable sensors, microfluidics, rehabilitative technology, and embedded and computer-aided design.

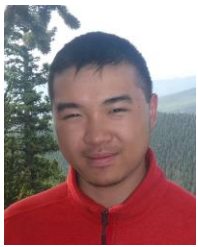

**Nam Tran** was born in Philadelphia, PA, USA in 1997. He received the B.S. degree in biomedical engineering with a minor in computer engineering from The George Washington University, in 2019.

From 2016 to 2019, he was an undergraduate research assistant in Dr. Zhenyu Li's laboratory at The George Washington University. He has also worked as a web developer for various organizations from 2014 until present.

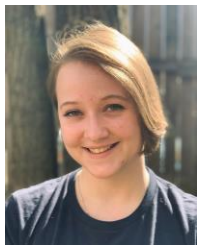

**Elizabeth Chorvinsky** was born in Chestertown, MD. She received a B.S. degree in Biological Sciences from the University of Maryland, College Park in 2014.

From 2015 to 2018, she worked as a research technician in the Program for Cell Enhancement and Technologies for Immunotherapies at Children's National Hospital, assisting in the development of immunotherapies for immunocompromised children.

Since late 2018, Elizabeth has been a member of the Airway Biology Research Group, also based at Children's National Hospital, where she applies her background in immunology to research focused on disorders of the respiratory system in children. In her budding scientific career, she has contributed to over a dozen published articles. Her research interests include steroid-resistant asthma, airway immunology and early childhood airway development at the molecular level.

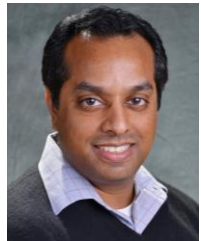

**Dinesh K. Pillai**, MD is a board-certified pediatrician and pediatric pulmonologist in the Division of Pulmonary Medicine at Children's National Hospital (CNH), an Associate Professor in the Departments of Pediatrics and Genomics & Precision Medicine at George Washington University (Washington, DC) and an investigator in the Center for Genetic Medicine Research (CNH). His basic science and translational research interest centers around airway epithelium related mechanisms involved in disease development and response to treatment for asthma. Dr. Pillai also serves as site Co-Investigator for the multicenter Inner-City Asthma Consortium. Clinically, he has developed a severe asthma clinic focused on treating children with refractory asthma. Overall, Dr. Pillai's work has led to the development of a comprehensive Severe Asthma Program at Children's National with a systems biology approach to treating and evaluating disease mechanisms in severe asthma.

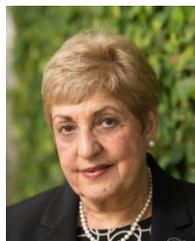

**Mona E. Zaghloul** (M'81–SM'85–F'80) received the M.A.Sc. degree in electrical engineering, the M. Math. degree in computer science and applied analysis, and the Ph.D. degree in electrical engineering from the University of Waterloo, Waterloo, ON, Canada, in 1970, 1971, and 1975, respectively.

Since 1984, she has been with the Semiconductor Devices Technology Division, National Institute of Standards and Technology, Gaithersburg, MD. She is also currently a Professor with the Department of Electrical and Computer Engineering, George Washington University, Washington, DC, where she was the Department Chair from 1994 to 1998 and currently directs the Institute of MEMS and VLSI Technology. The Institute of MEMS and VLSI Technology encompasses several interdisciplinary faculties and over a dozen graduate students with funding from several agencies. She is working in the research area of Sensors, Biosensors, MEMS applications to sensors with Integrated Electronics, Flexible and Wearable sensors, Nano-sensors and its applications to biomedical health. She published more than 110 Journal papers, and over 300 conferences papers, 3 books and 5 book chapters. She is the associate editor for IEEE-BIOCAS Journal, where she is responsible for Biosensors publications.

Dr. Zaghloul was a recipient of the 50th year Gold Jubilee Medal from the IEEE Circuits and Systems Society in recognition for her outstanding contribution to the society. She was the Vice President of the IEEE-CAS Technical Activities from 1999 to 2001. She was the IEEE Sensors Council President (2009-2010). She received Honorary Doctor Degree from University Waterloo in 2007 in recognition of her academic career in the international electrical engineering community, presented to her at the celebration of the University's 50th anniversary.

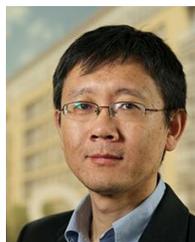

**Zhenyu Li** (M '08) received the B.S. degree in precision instrument from Tsinghua University, China, in 1999, the M.S degree in electrical engineering from University of California at Santa Barbara, CA, in 2000, and the Ph.D. degree in electrical engineering from the California Institute of Technology (Caltech), Pasadena, CA, in 2008.

From 2008 to 2010, he was a postdoctoral scholar at the Howard Hughes Medical Institute Janelia Research Campus and Caltech. Now he is Associate Professor with the Biomedical Engineering Department, The George Washington University, Washington DC. His current research interests include the development of biosensors and medical devices using microfluidics, MEMS, Optofluidics and flexible electronics. He has published over 50 peer-reviewed articles and holds seven patents.